\documentclass[orivec,a4paper,10pt]{llncs}

\usepackage{amsfonts}
\usepackage{amsmath}

\usepackage{algorithm}
\usepackage[noend]{algpseudocode}

\usepackage{tikz}
\usetikzlibrary{calc,shapes,arrows,patterns,automata,positioning}
\usetikzlibrary{decorations.pathreplacing}

\usepackage{thm-restate}
\usepackage{xspace}
\usepackage{cite}
\usepackage{paralist}

\usepackage{microtype}

\usepackage{dcolumn}
\newcolumntype{P}[1]{D{(}{\,(}{#1}}
\newcolumntype{+}[1]{D{+}{{}+}{#1}}
\newcommand{\hhline}{\hline\\[-3mm]}

\newenvironment{myproof}{\begin{proof}}{\qed\end{proof}}
\newenvironment{myexample}{\begin{example}}{\hfill$\lozenge$\end{example}}

\newcommand{\goal}{GOAL\xspace}
\newcommand{\rabit}{RABIT\xspace}
\newcommand{\roll}{ROLL\xspace}
\newcommand{\rollH}{ROLL${}_{H}$\xspace}
\newcommand{\rollB}{ROLL${}_{B}$\xspace}
\newcommand{\spot}{SPOT\xspace}
\newcommand{\buechi}{B\"uchi\xspace}

\newcommand{\naturals}{\mathbb{N}}

\newcommand{\appear}[2]{\#(#1, #2)}
\newcommand{\tree}{\mathcal{T}}
\newcommand{\treenode}[1]{\langle #1 \rangle}

\newcommand{\setcond}[2]{\{\, #1 \mid #2 \,\}}
\newcommand{\setnocond}[1]{\{#1\}}

\newcommand{\mc}{\mathsf{MC}^{2}}
\newcommand{\imc}{\mathsf{IMC}^{2}}
\newcommand{\prob}{\mathbf{Pr}}
\newcommand{\bigO}{\mathcal{O}}

\newcommand{\infset}[2][]{\mathop{\mathrm{Inf}}(#1{#2})}

\renewcommand{\phi}{\varphi}
\renewcommand{\epsilon}{\varepsilon}

\newcommand{\word}{\alpha}
\newcommand{\langSymbol}{\mathcal{L}}
\newcommand{\lang}[1]{\langSymbol(#1)}

\newcommand{\omegaRegLang}{L}
\newcommand{\size}[1]{|#1|}
\newcommand{\UP}[1]{\text{UP}(#1)}

\newcommand{\alphabet}{\Sigma}
\newcommand{\emptyword}{\lambda}
\newcommand{\finwords}{\alphabet^{*}}
\newcommand{\infwords}{\alphabet^{\omega}}
\newcommand{\poswords}{\alphabet^{+}}

\newcommand{\complementSymbol}{\mathsf{c}}
\newcommand{\complement}[1]{{#1}^{\complementSymbol}}

\newcommand{\transtyle}[1]{\mathrm{#1}}

\newcommand{\baut}[1][B]{\mathcal{#1}}
\newcommand{\astates}{Q}
\newcommand{\ainit}{\astates_{I}}

\newcommand{\amat}{\transtyle{T}}
\newcommand{\afinal}{\astates_{F}}

\newcommand{\arun}{\rho}
\newcommand{\afrun}{\sigma}

\newcommand{\terminatingLasso}[1][\afrun]{#1\bot}

\newcommand{\confprob}{\delta}
\newcommand{\errorprob}{\epsilon}
\newcommand{\ceprob}{p_{Z}}

\newcommand{\stopprob}{p_{\bot}}

\allowdisplaybreaks

\begin{document}

\title{Proving Non-Inclusion of \buechi Automata \\
 based on Monte Carlo Sampling\thanks{This work has been supported by the Guangdong Science and Technology Department (grant no.\ 2018B010107004) and by the National Natural Science Foundation of China (grant nos.\ 61761136011, 61532019).}}

\author{Yong Li\inst{1}, Andrea Turrini\inst{1,3}, Xuechao Sun\inst{1,2}, Lijun Zhang\inst{1,2,3}}
\institute{
State Key Laboratory of Computer Science, \\
Institute of Software, Chinese Academy of Sciences, Beijing, China
\and
University of Chinese Academy of Sciences, Beijing, China
\and
Institute of Intelligent Software, Guangzhou, China
}
\maketitle
\thispagestyle{plain}

\begin{abstract}
The search for a proof of correctness and the search for counterexamples (bugs) are complementary aspects of verification.
In order to maximize the practical use of verification tools it is better to pursue them at the same time.
While this is well-understood in the termination analysis of programs, this is not the case for the language inclusion analysis of \buechi automata, where research mainly focused on improving algorithms for proving language inclusion, with the search for counterexamples left to the expensive complementation operation.

In this paper, we present $\imc$, a specific algorithm for proving \buechi automata non-inclusion $\lang{\baut[A]} \not\subseteq \lang{\baut[B]}$, based on Grosu and Smolka's algorithm $\mc$ developed for Monte Carlo model checking against LTL formulas.
The algorithm we propose takes $M = \lceil \ln \delta / \ln (1-\epsilon) \rceil$ random lasso-shaped samples from $\baut[A]$ to decide whether to reject the hypothesis $\lang{\baut[A]} \not\subseteq \lang{\baut[B]}$, for given error probability $\epsilon$ and confidence level $1 - \delta$.
With such a number of samples, $\imc$ ensures that the probability of witnessing $\lang{\baut[A]} \not\subseteq \lang{\baut[B]}$ via further sampling is less than $\delta$, under the assumption that the probability of finding a lasso counterexample is larger than $\epsilon$.
Extensive experimental evaluation shows that $\imc$ is a fast and reliable way to find counterexamples to \buechi automata inclusion.
\end{abstract}

\setcounter{footnote}{0}

\section{Introduction}
\label{sec:intro}

The language inclusion checking of \buechi automata is a fundamental problem in the field of automated verification.
Specially, in the automata-based model checking~\cite{DBLP:conf/lics/VardiW86} framework, when both system and specification are given as \buechi automata, the model checking problem of verifying whether some system's behavior violates the specification reduces to a language inclusion problem between the corresponding \buechi automata.

In this paper, we target at the language inclusion checking problem of \buechi automata.
Since this problem has already been proved to be PSPACE-complete~\cite{DBLP:conf/cav/KupfermanV96a}, researchers have been focusing on devising algorithms to reduce its practical cost.
A na\"{i}ve approach to checking the inclusion between \buechi automata $\baut[A]$ and $\baut[B]$ is to first construct a complement automaton $\complement{\baut[B]}$ such that $\lang{\complement{\baut[B]}} = \infwords \setminus \lang{\baut[B]}$ and then to check the language emptiness of $\lang{\baut[A]} \cap \lang{\complement{\baut[B]}}$, which is the algorithm implemented in \spot~\cite{DBLP:conf/atva/Duret-LutzLFMRX16}, a highly optimized symbolic tool for manipulating LTL formulas and $\omega$-automata.

The bottleneck of this approach is computing the automaton $\complement{\baut[B]}$, which can be exponentially larger than $\baut[B]$~\cite{DBLP:journals/lmcs/Yan08}.
As a result,  various optimizations ---such as \emph{subsumption} and \emph{simulation}--- 
have been proposed to avoid exploring the whole state-space of $\complement{\baut[B]}$, see, e.g.,~\cite{DBLP:journals/corr/abs-0902-3958,DBLP:conf/tacas/FogartyV10,DBLP:conf/cav/AbdullaCCHHMV10,DBLP:conf/concur/AbdullaCCHHMV11,DBLP:journals/lmcs/ClementeM19,DBLP:journals/siamcomp/EtessamiWS05}.
For instance, \rabit is currently the state-of-the-art tool for checking language inclusion between \buechi automata, which has integrated the simulation and subsumption techniques proposed in~\cite{DBLP:conf/cav/AbdullaCCHHMV10,DBLP:conf/concur/AbdullaCCHHMV11,DBLP:journals/lmcs/ClementeM19}.
All these techniques improving the language inclusion checking, however, focus on \emph{proving} inclusion.
In particular, the simulation techniques in~\cite{DBLP:journals/siamcomp/EtessamiWS05,DBLP:journals/lmcs/ClementeM19} are specialized algorithms mainly proposed to obtain such proof, which ensures that for every initial state $q_{a}$ of $\baut[A]$, there is an initial state $q_{b}$ of $\baut[B]$ that simulates every possible behavior from $q_{a}$.

From a practical point of view, it is widely believed that the witness of a counterexample (or bug) found by a verification tool is equally valuable as a proof for the correctness of a program;
we would argue that showing why a program violates the specification is also intuitive for a programmer, since it gives a clear way to identify and correct the error.
Thus, the search for a proof and the search for counterexamples (bugs) are complementary activities that need to be pursued at the same time in order to maximize the practical use of verification tools.
This is well-understood in the termination analysis of programs, as the techniques for searching the proof of the termination~\cite{DBLP:journals/corr/LeikeH15,DBLP:conf/icalp/BradleyMS05,DBLP:conf/cav/Ben-AmramG17} and the counterexamples~\cite{DBLP:conf/popl/GuptaHMRX08, DBLP:conf/tacas/LeikeH18,DBLP:conf/cade/EmmesEG12} are evolving concurrently.
Counterexamples to \buechi automata language inclusion, instead, are the byproducts of a failure while proving language inclusion.
Such a failure may be recognized after a considerable amount of efforts has been spent on proving inclusion, in particular when the proposed improvements are not effective.
In this work, instead, we focus directly on the problem of finding a counterexample to language inclusion.

The main contribution is a novel algorithm called $\imc$ for showing language non-inclusion based on sampling and statistical hypothesis testing. 
Our algorithm is inspired by the Monte Carlo approach proposed in~\cite{DBLP:conf/tacas/GrosuS05} for model checking systems against LTL specifications.
The algorithm proposed in~\cite{DBLP:conf/tacas/GrosuS05} takes as input a \buechi automaton $\baut[A]$ as system and an LTL formula $\phi$ as specification and then checks whether $\baut[A] \not\models \phi$ by equivalently checking $\lang{\baut[A]} \not\subseteq \lang{\baut[B]_{\phi}}$, where $\baut[B]_{\phi}$ is the \buechi automaton constructed for $\phi$.
The main idea of the algorithm for showing $\lang{\baut[A]} \not\subseteq \lang{\baut[B]_{\phi}}$ is to sample lasso words from the product automaton $\baut[A] \times \baut[B]^{\complementSymbol}_{\phi}$ for $\lang{\baut[A]} \cap \lang{\baut[B]^{\complementSymbol}_{\phi}}$; 
lasso words are of the form $uv^{\omega}$ and are obtained as soon as a state is visited twice. 
If one of such lasso words is accepted by $\baut[A] \times \baut[B]^{\complementSymbol}_{\phi}$, then it is surely a witness to $\lang{\baut[A]} \not\subseteq \lang{\baut[B]_{\phi}}$, i.e., a counterexample to $\baut[A] \models \phi$.
Since in~\cite{DBLP:conf/tacas/GrosuS05} the algorithm gets an LTL formula $\phi$ as input, the construction of $\baut[B]^{\complementSymbol}_{\phi}$ reduces to the construction of $\baut[B]_{\neg\phi}$ and it is widely assumed that the translation into a \buechi automaton is equally efficient for a formula and its negation. 
In this paper, we consider the general case, namely the specification is given as a generic \buechi automaton $\baut$, where the construction of $\complement{\baut}$ from $\baut$ can be very expensive~\cite{DBLP:journals/lmcs/Yan08}.

To avoid the heavy generation of $\complement{\baut[B]}$, the algorithm $\imc$ we propose directly sampling lasso words in $\baut[A]$, without making the product $\baut[A] \times \complement{\baut[B]}$.
We show that usual lasso words, like the ones used in~\cite{DBLP:conf/tacas/GrosuS05}, do not suffice in our case, and propose a rather intriguing sampling procedure.
We allow the lasso word $uv^{\omega}$ to visit each state of $\baut[A]$ multiple times, i.e., the run $\afrun$ of $\baut[A]$ on the finite word $uv$ can present small cycles on both the $u$ and the $v$ part of the lasso word.
We achieve this by setting a bound $K$ on the number of times a state can be visited:
each state in $\afrun$ is visited at most $K-1$ times, except for the last state of $\afrun$ that is visited at most $K$ times.
We show that $\imc$ gives a probabilistic guarantee in terms of finding a counterexample to inclusion when $K$ is sufficiently large, as described in Theorem~\ref{thm:imc-complexity}.
This notion of generalized lasso allows our approach to find counterexamples that are not valid lassos in the usual sense.
The extensive experimental evaluation  shows that our approach is generally very fast and reliable in finding counterexamples to language inclusion.
In particular, the prototype tool we developed is able to manage easily \buechi automata with very large state space and alphabet on which the state-of-the-art tools such as \rabit and \spot fail.
This makes our approach fit very well among tools that make use of \buechi automata language inclusion tests, since it can quickly provide counterexamples before having to rely on the possibly time and resource consuming structural methods, in case an absolute guarantee about the result of the inclusion test is desired.

\paragraph*{Organization of the paper.} 
In the remainder of this paper, we briefly recall some known results about \buechi automata in Section~\ref{sec:preliminaries}.
We then present the algorithm $\imc$ in Section~\ref{sec:algorithm} and give the experimental results in Section~\ref{sec:experiments} before concluding the paper with some remark in Section~\ref{sec:conclusion}.

\section{Preliminaries}
\label{sec:preliminaries}

\subsubsection{\buechi Automata}
\label{ssec:preliminariesBuechiAutomata}

Let $\alphabet$ be a finite set of \emph{letters} called \emph{alphabet}. A finite sequence of letters is called a \emph{word}.
An infinite sequence of letters is called an \emph{$\omega$-word}.
We use $\size{\word}$ to denote the length of the finite word $\word$ and we use $\emptyword$ to represent the empty word, i.e., the word of length $0$.
The set of all finite words on $\alphabet$ is denoted by $\finwords$, and the set of all $\omega$-words is denoted by $\infwords$.
Moreover, we also denote by $\poswords$ the set $\finwords \setminus \setnocond{\emptyword}$.

A \emph{nondeterministic \buechi automaton} (NBA) is a tuple $\baut = (\alphabet, \astates, \ainit, \amat, \afinal)$, consisting of
a finite \emph{alphabet} $\alphabet$ of input letters,
a finite set $\astates$ of \emph{states} with a non-empty set $\ainit \subseteq \astates$ of \emph{initial states},
a set $\amat \subseteq \astates \times \alphabet \times \astates$ of \emph{transitions},
and
a set $\afinal \subseteq \astates$ of \emph{accepting states}.

A \emph{run} of an NBA $\baut$ over an $\omega$-word $\word = a_{0} a_{1} a_{2} \cdots \in \infwords$ is an infinite alternation of states and letters $\arun = q_{0} a_{0} q_{1} a_{1} q_{2} \cdots \in (\astates \times \alphabet) ^{\omega}$ such that $q_{0} \in \ainit$ and, for each $i \geq 0$, $\big(\arun(i), a_{i}, \arun(i+1)\big) \in \amat$ where $\arun(i) = q_{i}$.
A run $\arun$ is \emph{accepting} if it contains infinitely many accepting states, i.e., $\infset{\arun} \cap \afinal \neq \emptyset$, where $\infset{\arun} = \setcond{q \in \astates}{\forall i \in \naturals. \exists j > i : \arun(j) = q}$.
An $\omega$-word $\word$ is \emph{accepted} by $\baut$ if $\baut$ has an accepting run on $\word$, and
the set of words $\lang{\baut} = \setcond{\word \in \infwords}{\text{$\word$ is accepted by $\baut$}}$ accepted by $\baut$ is called its \emph{language}.

We call a subset of $\infwords$ an \emph{$\omega$-language} and the language of an NBA an \emph{$\omega$-regular language}.
Words of the form $uv^{\omega}$ are called \emph{ultimately periodic} words.
We use a pair of finite words $(u,v)$ to denote the ultimately periodic word $w = uv^{\omega}$.
We also call $(u, v)$ a \emph{decomposition} of $w$.
For an $\omega$-language $\omegaRegLang$, let $\UP{\omegaRegLang} = \setcond{uv^{\omega} \in \omegaRegLang}{u \in \finwords, v \in \poswords}$ be the set of all ultimately periodic words in $\omegaRegLang$.
The set of ultimately periodic words can be seen as the fingerprint of $\omegaRegLang$:
\begin{theorem}[Ultimately Periodic Words~\cite{Buchi1990}]
\label{thm:ultimately-eq-words}
	Let $\omegaRegLang$, $\omegaRegLang'$ be two $\omega$-regular languages.
	Then $\omegaRegLang = \omegaRegLang'$ if, and only if, $\UP{\omegaRegLang} = \UP{\omegaRegLang'}$.
\end{theorem}
An immediate consequence of Theorem~\ref{thm:ultimately-eq-words} is that, for any two $\omega$-regular languages $\omegaRegLang_{1}$ and $\omegaRegLang_{2}$,
if $\omegaRegLang_{1} \neq \omegaRegLang_{2}$ then there is an ultimately periodic word $xy^{\omega} \in \big(\UP{\omegaRegLang_{1}} \setminus \UP{\omegaRegLang_{2}}\big) \cup \big(\UP{\omegaRegLang_{2}} \setminus \UP{\omegaRegLang_{1}}\big)$.
It follows that $xy^{\omega} \in \omegaRegLang_{1} \setminus \omegaRegLang_{2}$ or $xy^{\omega} \in \omegaRegLang_{2} \setminus \omegaRegLang_{1}$.
Let $\baut[A]$, $\baut[B]$ be two NBAs and assume that $\lang{\baut[A]} \setminus \lang{\baut[B]} \neq \emptyset$.
One can find an ultimately periodic word $xy^{\omega} \in \lang{\baut[A]} \setminus \lang{\baut[B]}$ as a counterexample to $\lang{\baut[A]} \subseteq \lang{\baut[B]}$.

Language inclusion between NBAs can be reduced to \emph{complementation}, \emph{intersection}, and \emph{emptiness} problems on NBAs.
The complementation operation of an NBA $\baut$ is to construct an NBA $\complement{\baut}$ accepting the complement language of $\lang{\baut}$, i.e., $\lang{\complement{\baut}} = \infwords \setminus \lang{\baut}$.

\begin{lemma}[cf.~\cite{DBLP:journals/tocl/KupfermanV01,DBLP:reference/mc/Kupferman18}]
\label{lem:buechiOperations}
	Let $\baut[A]$, $\baut[B]$ be NBAs with $n_{a}$ and $n_{b}$ states, respectively.
	\begin{enumerate}
	\item
	\label{itm:lem:buechiOperations:complement}
		It is possible to construct an NBA $\complement{\baut[B]}$ such that $\lang{\complement{\baut[B]}} = \infwords \setminus \lang{\baut[B]}$ whose number of states is at most $(2n_{b}+2)^{n_{b}} \times 2^{n_{b}}$, by means of the complement construction.
	\item
	\label{itm:lem:buechiOperations:intersection}
		It is possible to construct an NBA $\baut[C]$ such that $\lang{\baut[C]} = \lang{\baut[A]} \cap \lang{\complement{\baut[B]}}$ whose number of states is at most $2 \times n_{a} \times (2n_{b}+2)^{n_{b}} \times 2^{n_{b}}$, by means of the product construction.
		Note that $\lang{\baut[A]} \subseteq \lang{\baut[B]}$ holds if and only if $\lang{\baut[C]} = \emptyset$ holds.
	\item
	\label{itm:lem:buechiOperations:emptiness}
		$\lang{\baut[C]} = \emptyset$ is decidable in time linear in the number of states of $\baut[C]$.
	\end{enumerate}
\end{lemma}
Further, testing whether an $\omega$-word $w$ is accepted by a \buechi automaton $\baut$ can be done in time polynomial in the size of the decomposition $(u,v)$ of $w = uv^{\omega}$.

\begin{lemma}[cf.~\cite{DBLP:reference/mc/Kupferman18}]
\label{lem:nba-membership}
	Let $\baut$ be an NBA with $n$ states and an ultimately periodic word $(u, v)$ with $\size{u} + \size{v} = m$.
	Then checking whether $uv^{\omega}$ is accepted by $\baut$ is decidable in time and space linear in $n \times m$.
\end{lemma}

\subsubsection{Random Sampling and Hypothesis Testing}
\label{ssec:pre-sampling-testing}

Statistical hypothesis testing is a statistical method to assign a confidence level to the correctness of the interpretation given to a small set of data sampled from a population, when this interpretation is extended to the whole population.

Let $Z$ be a Bernoulli random variable and $X$ the random variable with parameter $p_{Z}$ whose value is the number of independent trials required until we see that $Z = 1$.
Let $\confprob$ be the \emph{significance level} that $Z = 1$ will not appear within $N$ trials.
Then $N = \lceil \ln \confprob / \ln(1 - \ceprob) \rceil$ is the number of attempts needed to get a counterexample with probability at most $1 - \confprob$.

If the exact value of $\ceprob$ is unknown, given an \emph{error probability} $\errorprob$ such that $\ceprob \geq \errorprob$, we have that $M = \lceil \ln \confprob / \ln(1 - \errorprob) \rceil \geq N = \lceil \ln \confprob / \ln(1 - \ceprob) \rceil$ ensures that $\ceprob \geq \errorprob \implies \prob[X \leq M] \geq 1 - \confprob$.
In other words, $M$ is the minimal number of attempts required to find a counterexample with probability $1 - \confprob$, under the assumption that $\ceprob \geq \errorprob$.
See, e.g.,~\cite{DBLP:conf/tacas/GrosuS05,Younes05} for more details about statistical hypothesis testing in the context of formal verification.

\section{Monte Carlo Sampling for Non-Inclusion Testing}
\label{sec:algorithm}

In this section we present our Monte Carlo sampling algorithm $\imc$ for testing non-inclusion between \buechi automata.

\subsection{$\mc$: Monte Carlo Sampling for LTL Model Checking}
\label{ssed:algorihtm:mc}

In~\cite{DBLP:conf/tacas/GrosuS05}, the authors proposed a Monte Carlo sampling algorithm $\mc$ for verifying whether a given system $A$ satisfies a Linear Temporal Logic (LTL) specification $\phi$. 
$\mc$ works directly on the product \buechi automaton $\baut[P]$ that accepts the language $\lang{A} \cap \lang{\baut[B]_{\neg \phi}}$. 
It essentially checks whether $\lang{\baut[P]}$ is empty.

First, $\mc$ takes two statistical parameters $\epsilon$ and $\sigma$ as input and computes the number of samples $M$ for this experiment.
Since every ultimately periodic word $xy^{\omega} \in \lang{\baut[P]}$ corresponds to some cycle run (or ``lasso'') in $\baut[P]$, $\mc$ can just find \emph{an accepting lasso} whose corresponding ultimately periodic word $xy^{\omega}$ is such that $xy^{\omega} \in \lang{\baut[P]}$.
In each sampling procedure, $\mc$ starts from a randomly chosen initial state and performs a random walk on $\baut[P]$'s transition graph until a state has been visited twice, which consequently gives a lasso in $\baut[P]$.
$\mc$ then checks whether there exists an accepting state in the repeated part of the sampled lasso.
If so, $\mc$ reports it as a counterexample to the verification, otherwise it continues with another sampling process if necessary.
The correctness of $\mc$ is straightforward, as the product automaton $\baut[P]$ is non-empty if and only if there is an accepting lasso.

\subsection{The Lasso Construction Fails for Language Inclusion}
\label{ssed:algorihtm:imc}

The Monte Carlo Sampling algorithm in~\cite{DBLP:conf/tacas/GrosuS05} operates directly on the product. 
For language inclusion, as discussed in the introduction, this is the bottleneck of the construction. 
Thus, we aim at a sampling algorithm operating on the automata $\baut[A]$ and $\baut[B]$, separately. 
With this in mind, we show first that, directly applying $\mc$ can be incomplete for language inclusion checking.

\begin{figure}[t]
	\centering
	\begin{tikzpicture}[->, >=stealth', node distance=1.5cm, auto, initial text={}]
	\path[use as bounding box] (-1,-1.15) rectangle (7.5,1.25);

		\begin{scope}
		\node[initial, state] (s1)      {$s_{1}$};
		\node[state, accepting] (s2) at ($(s1) + (2,0)$)    {$s_{2}$};
		\node at ($(s1) + (1, -1)$) {$\baut[A]$};

		\path (s1) edge [loop above] node {$a$} (s1)
					edge node {$b$}              (s2)
				(s2) edge [loop above] node {$b$} (s2);
		\end{scope}

		\begin{scope}[xshift=5cm]
		\node[initial, state] (q1)      {$q_{1}$};
		\node[state, accepting] (q2) at ($(q1) + (2,0)$)    {$q_{2}$};
		\node at ($(q1) + (1, -1)$) {$\baut[B]$};

		\path (q1) edge node {$b$} (q2)
				(q2) edge [loop above] node {$b$} (q2);
		\end{scope}
	\end{tikzpicture}
	\caption{Two NBAs $\baut[A]$ and $\baut[B]$.}
	\label{fig:buechi-example}
\end{figure}

\begin{myexample}
	Consider checking the language inclusion of the \buechi automata $\baut[A]$ and $\baut[B]$ in Fig.~\ref{fig:buechi-example}.
	As we want to exploit $\mc$ to find a counterexample to the inclusion, we need to sample a word from $\baut[A]$ that is accepted by $\baut[A]$ but not accepted by $\baut[B]$.
	In~\cite{DBLP:conf/tacas/GrosuS05}, the sampling procedure is terminated as soon as a state is visited twice.
	Thus, the set of lassos that can be sampled by $\mc$ is $\setnocond{s_{1} a s_{1}, s_{1} b s_{2} b s_{2}}$, which yields the set of words $\setnocond{a^{\omega}, b^{\omega}}$.
	It is easy to see that neither of these two words is a counterexample to the inclusion.
	The inclusion, however, does not hold:
	the word $ab^{\omega} \in \lang{\baut[A]} \setminus \lang{\baut[B]}$ is a counterexample.
\end{myexample}

According to Theorem~\ref{thm:ultimately-eq-words}, if $\lang{\baut[A]} \setminus\lang{\baut[B]} \neq \emptyset$, then there must be an ultimately periodic word $xy^{\omega} \in \lang{\baut[A]} \setminus \lang{\baut[B]}$ as a counterexample to the inclusion.
It follows that there exists some lasso in $\baut[A]$ whose corresponding ultimately periodic word is a counterexample to the inclusion.
The \emph{limit of $\mc$ in checking the inclusion} is that $\mc$ only samples simple lasso runs, which may miss non-trivial lassos in $\baut[A]$ that correspond to counterexamples to the inclusion. 
The reason that it is sufficient for checking non-emptiness in the product automaton is due to the fact that the product automaton already synchronizes behaviors of $\baut[A]$ and $\baut[B]_{\neg\phi}$.

In the remainder of this section, we shall propose a new  definition of lassos by allowing multiple occurrences of states, which is the key point of our extension.

\subsection{$\imc$: Monte Carlo Sampling for Inclusion Checking}
We now present our Monte Carlo sampling algorithm called $\imc$ specialized for testing the language inclusion between two given NBAs $\baut[A]$ and $\baut[B]$.

We first define the lassos of $\baut[A]$ in Definition~\ref{def:lasso} and show how to compute the probability of a sample lasso in Definition~\ref{def:lasso-prob}.
Then we prove that with our definition of the lasso probability space in $\baut[A]$, the probability of a sample lasso whose corresponding ultimately periodic word $xy^{\omega}$ is a counterexample to the inclusion is greater than $0$ under the hypothesis $\lang{\baut[A]} \not\subseteq \lang{\baut[B]}$.
Thus we eventually get for sure a sample from $\baut[A]$ that is a counterexample to the inclusion, if inclusion does not hold.
In other words, we are able to obtain a counterexample to the inclusion with high probability from a large amount of samples.

In practice, a lasso of $\baut[A]$ is sampled via a random walk on $\baut[A]$'s transition graph, starting from a randomly chosen initial state and picking uniformly one outgoing transition.
In the following, we fix a natural number $K \geq 2$ unless explicitly stated otherwise and two NBAs $\baut[A] = (\alphabet, \astates, \ainit, \amat, \afinal)$ and $\baut[B]$.
We assume that each state in $\baut[A]$ can reach an accepting state and has at least one outgoing transition.
Note that each NBA $\baut[A]$ with $\lang{\baut[A]} \neq \emptyset$ can be pruned to satisfy such assumption;
only NBAs $\baut[A]'$ with $\lang{\baut[A]'} = \emptyset$ do not satisfy the assumption, but for these automata the problem $\lang{\baut[A]'} \subseteq \lang{\baut[B]}$ is trivial.

\begin{definition}[Lasso]
\label{def:lasso}
	Given two NBAs $\baut[A]$, $\baut[B]$ and a natural $K \geq 2$, a finite run $\afrun = q_{0} a_{0} q_{1} \cdots a_{n-1} q_{n} a_{n} q_{n+1}$ of $\baut[A]$ is called a \emph{$K$-lasso} if
	\begin{inparaenum}[(1)]
	\item
		each state in $\setnocond{q_{0}, \dotsc, q_{n}}$ occurs at most $K - 1$ times in $q_{0} a_{0} q_{1} \cdots a_{n-1} q_{n}$ and
	\item
		$q_{n+1} = q_{i}$ for some $0 \leq i \leq n$ (thus, $q_{n+1}$ occurs at most $K$ times in $\afrun$).
	\end{inparaenum}
	We write $\terminatingLasso$ for the \emph{terminating} $K$-lasso $\afrun$, where $\bot$ is a fresh symbol denoting termination.
	We denote by $S^{K}_{\baut[A]}$ the set of all terminating $K$-lassos for $\baut[A]$.

	We call $\terminatingLasso \in S^{K}_{\baut[A]}$ a \emph{witness} for $\lang{\baut[A]} \setminus\lang{\baut[B]} \neq \emptyset$ if the associated $\omega$-word $(a_{0} \cdots a_{i-1}, a_{i} \cdots a_{n})$ is accepted by $\baut[A]$ but not accepted by $\baut[B]$.
\end{definition}
It is worth noting that not every finite cyclic run of $\baut[A]$ is a valid $K$-lasso.
Consider the NBA $\baut[A]$ shown in Fig.~\ref{fig:buechi-example} for instance:
the run $s_{1} a s_{1} b s_{2} b s_{2}$ is not a lasso when $K = 2$ since by Definition~\ref{def:lasso} every state except the last one is allowed to occur at most $K - 1 = 1$ times;
$s_{1}$ clearly violates this requirement since it occurs twice and it is not the last state of the run.
The run $s_{1} b s_{2} b s_{2}$ instead is obviously a valid lasso when $K = 2$.

\begin{remark}
\label{rem:KlassoProperties}
	A $K$-lasso $\afrun$ is also a $K'$-lasso for any $K' > K$.
	Moreover, a terminating $K$-lasso can be a witness without being an accepting run:
	according to Definition~\ref{def:lasso}, a terminating $K$-lasso $\terminatingLasso$ is a witness if its corresponding word $uv^{\omega}$ is accepted by $\baut[A]$ but not accepted by $\baut[B]$.
	This does not imply that $\afrun$ is an accepting run, since there may be another run $\afrun'$ on the same word $uv^{\omega}$ that is accepting.
\end{remark}

In order to define a probability space over $S^{K}_{\baut[A]}$, we first define the probability of a terminating $K$-lasso of $\baut[A]$.
We denote by $\appear{\afrun}{q}$ the number of occurrences of the state $q$ in the $K$-lasso $\afrun$.

\begin{definition}[Lasso Probability]
\label{def:lasso-prob}
	Given an NBA $\baut[A]$, a natural number $K \geq 2$, and a stopping probability $\stopprob \in (0,1)$, the probability $\prob_{\stopprob}[\terminatingLasso]$ of a terminating $K$-lasso $\terminatingLasso = \terminatingLasso[q_{0} a_{0} \cdots q_{n} a_{n} q_{n+1}] \in S^{K}_{\baut[A]}$ is defined as follows:
	\begin{align*}
		&\prob_{\stopprob}[\terminatingLasso] =
			\begin{cases}
				\prob'_{\stopprob}[\afrun] & \text{if $\appear{\afrun}{q_{n+1}} = K$,}\\				
				\stopprob \cdot \prob'_{\stopprob}[\afrun] & \text{if $\appear{\afrun}{q_{n+1}} < K$;}
			\end{cases}
			\\
		&\prob'_{\stopprob}[\afrun] =
			\begin{cases}
				\frac{1}{\size{\ainit}} & \text{if $\afrun = q_{0}$;}\\
				\prob'_{\stopprob}[\afrun'] \cdot \pi[q_{l} a_{l} q_{l+1}] & \text{if $\afrun = \afrun' a_{l} q_{l+1}$ and $\appear{\afrun'}{q_{l}} = 1$;}\\
				(1 {-} \stopprob) \cdot \prob'_{\stopprob}[\afrun'] \cdot \pi[q_{l} a_{l} q_{l+1}] &  \text{if $\afrun = \afrun' a_{l} q_{l+1}$ and $\appear{\afrun'}{q_{l}} > 1$,}
			\end{cases}
	\end{align*}
	where $\pi[q a q'] = \frac{1}{m} $ if $(q, a, q') \in \amat$ and $\size{\amat(q)} = m$, $0$ otherwise.
	
	We extend $\prob_{\stopprob}$ to sets of terminating $K$-lassos in the natural way, i.e., for $S \subseteq S^{K}_{\baut[A]}$, $\prob_{\stopprob}[S] = \sum_{\terminatingLasso \in S} \prob_{\stopprob}[\terminatingLasso]$.
\end{definition}
Assume that the current state of run $\afrun$ is $q$.
Intuitively, if the last state $s$ of the run $\afrun$ has been already visited at least twice but less than $K$ times, the run $\afrun$ can either terminate at $s$ with probability $\stopprob$ or continue with probability $1 - \stopprob$ by taking uniformly one of the outgoing transitions from the state $q$.
However, as soon as the state $q$ has been visited $K$ times, the run $\afrun$ has to terminate.

\begin{figure}[t]
	\centering
	\resizebox{\linewidth}{!}{
	\begin{tikzpicture}[->, >=stealth', shorten >=0pt, auto]
		\path[use as bounding box] (-6.05,0.1) rectangle (6.75,-5.3);
		
		\node (root) at (0,0) {$\treenode{\emptyword, 1}$};
		\node (s1) at ($(root) + (0,-1.05)$) {$\treenode{s_{1},1}$};
		\node (s1bs2) at ($(s1) + (2.5,-1.05)$) {$\treenode{s_{1} b s_{2}, \frac{1}{2}}$};
		\node (s1bs2bs2) at ($(s1bs2) + (0,-1.05)$) {$\treenode{s_{1} b s_{2} b s_{2}, \frac{1}{2}}$};
		\node (s1bs2bs2h) at ($(s1bs2bs2) + (3,0)$) {$\treenode{\terminatingLasso[s_{1} b s_{2} b s_{2}], \frac{1}{4}}$};
		\node (s1bs2bs2bs2) at ($(s1bs2bs2) + (0,-1.05)$) {$\treenode{s_{1} b s_{2} b s_{2} b s_{2}, \frac{1}{4}}$};
		\node (s1bs2bs2bs2h) at ($(s1bs2bs2bs2) + (3,0)$) {$\treenode{\terminatingLasso[s_{1} b s_{2} b s_{2} b s_{2}], \frac{1}{4}}$};
		\node (s1as1) at ($(s1) + (-2.5,-1.05)$) {$\treenode{s_{1} a s_{1}, \frac{1}{2}}$};
		\node (s1as1h) at ($(s1as1) + (-2.5,0)$) {$\treenode{\terminatingLasso[s_{1} a s_{1}], \frac{1}{4}}$};
		\node (s1as1as1) at ($(s1as1) + (0,-1.05)$) {$\treenode{s_{1} a s_{1} a s_{1}, \frac{1}{8}}$};
		\node (s1as1as1h) at ($(s1as1as1) + (-2.5,0)$) {$\treenode{\terminatingLasso[s_{1} a s_{1} a s_{1}], \frac{1}{8}}$};
		\node (s1as1bs2) at ($(s1as1) + (2.5,-1.05)$) {$\treenode{s_{1} a s_{1} b s_{2}, \frac{1}{8}}$};
		\node (s1as1bs2bs2) at ($(s1as1bs2) + (0,-1.05)$) {$\treenode{s_{1} a s_{1} b s_{2} b s_{2}, \frac{1}{8}}$};
		\node (s1as1bs2bs2h) at ($(s1as1bs2bs2) + (-4,0)$) {$\treenode{\terminatingLasso[s_{1} a s_{1} b s_{2} b s_{2}], \frac{1}{16}}$};
		\node (s1as1bs2bs2bs2) at ($(s1as1bs2bs2) + (0,-1.05)$) {$\treenode{s_{1} a s_{1} b s_{2} b s_{2} b s_{2}, \frac{1}{16}}$};
		\node (s1as1bs2bs2bs2h) at ($(s1as1bs2bs2bs2) + (-4,0)$) {$\treenode{\terminatingLasso[s_{1} a s_{1} b s_{2} b s_{2} b s_{2}], \frac{1}{16}}$};

		\draw (root) to (s1);
		\draw (s1) to (s1bs2);
		\draw (s1bs2) to (s1bs2bs2);
		\draw (s1bs2) to (s1bs2bs2);
		\draw (s1bs2bs2) to (s1bs2bs2h);
		\draw (s1bs2bs2) to (s1bs2bs2bs2);
		\draw (s1bs2bs2bs2) to (s1bs2bs2bs2h);
		\draw (s1) to (s1as1);
		\draw (s1as1) to (s1as1h);
		\draw (s1as1) to (s1as1as1);
		\draw (s1as1as1) to (s1as1as1h);
		\draw (s1as1) to (s1as1bs2);
		\draw (s1as1bs2) to (s1as1bs2bs2);
		\draw (s1as1bs2bs2) to (s1as1bs2bs2h);
		\draw (s1as1bs2bs2) to (s1as1bs2bs2bs2);
		\draw (s1as1bs2bs2bs2) to (s1as1bs2bs2bs2h);
	\end{tikzpicture}
	}
	\caption{An instance $\tree$ of the trees used in the proof of Theorem~\ref{thm:discrete-distribution-inclusion}. Each leaf node is labeled with a terminating $3$-lasso $\terminatingLasso \in S^{3}_{\baut[A], \baut[B]}$ for the NBAs $\baut[A]$ and $\baut[B]$ shown in Fig.~\ref{fig:buechi-example}, and its corresponding probability value $\prob_{\frac{1}{2}}[\terminatingLasso]$.}
	\label{fig:lasso-prob-space-tree-example}
\end{figure}

\begin{restatable}[Lasso Probability Space]{theorem}{lassoprobspace}
\label{thm:discrete-distribution-inclusion}
	Let $\baut[A]$ be an NBA, $K \geq 2$, and a stopping probability $\stopprob \in (0,1)$.
	The $\sigma$-field $(S^{K}_{\baut[A]}, 2^{S^{K}_{\baut[A]}})$ together with $\prob_{\stopprob}$ defines a discrete probability space.
\end{restatable}
\begin{myproof}[sketch]
	The facts that $\prob_{\stopprob}[\afrun]$ is a non-negative real value for each $\afrun \in S$ and that $\prob_{\stopprob}[S_{1} \cup S_{2}] = \prob_{\stopprob}[S_{1}] + \prob_{\stopprob}[S_{2}]$ for each $S_{1}, S_{2} \subseteq S^{K}_{\baut[A]}$ such that $S_{1} \cap S_{2} = \emptyset$ are both immediate consequences of the definition of $\prob_{\stopprob}$.
	
	The interesting part of the proof is about showing that $\prob_{\stopprob}[S^{K}_{\baut[A]}] = 1$.
	To prove this, we make use of a tree $\tree = (N, \treenode{\emptyword, 1}, E)$, like the one shown in Fig.~\ref{fig:lasso-prob-space-tree-example}, whose nodes are labelled with finite runs and probability values.
	In particular, we label the leaf nodes of $\tree$ with the terminating $K$-lassos in $S^{K}_{\baut[A]}$ while we use their finite run prefixes to label the internal nodes.
	Formally, the tree $\tree$ is constructed as follows.
	Let $P = \setcond{\afrun' \in \astates \times (\alphabet \times \astates)^{*}}{\text{$\afrun'$ is a prefix of some $\terminatingLasso \in S^{K}_{\baut[A]}$}}$ be the set of prefixes of the $K$-lassos in $S^{K}_{\baut[A]}$.
	$\tree$'s components are defined as follows.
	\begin{itemize}
	\item
		$N = \big(P \times (0, 1]\big) \cup \big(S^{K}_{\baut[A]} \times (0, 1]\big) \cup \setnocond{\treenode{\emptyword, 1}}$ is the set of nodes,
	\item
		$\treenode{\emptyword, 1}$ is the root of the tree, and
	\item
		$E \subseteq \Big(\setnocond{\treenode{\emptyword, 1}} \times \big(P \times (0, 1]\big)\Big) \cup \Big(P \times (0, 1]\Big)^{2} \cup \Big(\big(P \times (0, 1]\big) \times \big(S^{K}_{\baut[A]} \times (0, 1]\big)\Big)$ is the set of edges defined as
		\begin{align*}
			E = & \hphantom{{} \cup {}}\setcond{(\treenode{\emptyword, 1}, \treenode{q, \frac{1}{\size{\ainit}}})}{q \in \ainit} \\
			& {} \cup \setcond{\big(\treenode{\afrun, p}, \treenode{\afrun a q, \frac{p}{\size{\amat(\afrun_{l})}}}\big)}{\afrun a q \in P \land \appear{\afrun}{\afrun_{l}} = 1} \\
			& {} \cup \setcond{\big(\treenode{\afrun, p}, \treenode{\afrun a q,\frac{p \cdot (1 - \stopprob)}{\size{\amat(\afrun_{l})}}}\big)}{\afrun a q \in P \land \appear{\afrun}{\afrun_{l}} > 1} \\
			& {} \cup \setcond{\big(\treenode{\afrun, p}, \treenode{\terminatingLasso, p}\big)}{\terminatingLasso \in S^{K}_{\baut[A]} \land \appear{\afrun}{\afrun_{l}} = K} \\
			& {} \cup \setcond{\big(\treenode{\afrun, p}, \treenode{\terminatingLasso, p \cdot \stopprob}\big)}{\terminatingLasso \in S^{K}_{\baut[A]} \land \appear{\afrun}{\afrun_{l}} < K}
		\end{align*}
		where $\afrun_{l}$ denotes the last state $s_{n}$ of the finite run $\afrun = s_{0} a_{0} s_{1} \dotsc a_{n-1} s_{n}$.
	\end{itemize}
	
	Then we show a correspondence between the reachable leaf nodes and the terminating $K$-lassos with their $\prob_{\stopprob}$ probability values, and that the probability value in each internal node equals the sum of the probabilities of its children.
	By the finiteness of the reachable part of the tree we then derive $\prob_{\stopprob}[S^{K}_{\baut[A]}] = 1$.
	
	The complete proof can be found in Appendix~\ref{app:proofs}.
\end{myproof}

\begin{myexample}[Probability of lassos]
\label{ex:prob-lassos}
	Consider the \buechi automaton $\baut[A]$ of Fig.~\ref{fig:buechi-example} and $\stopprob = \frac{1}{2}$.
	For $K = 2$, there are only two terminating $2$-lassos, namely $\terminatingLasso[s_{1} a s_{1}]$ and $\terminatingLasso[s_{1} b s_{2} b s_{2}]$.
	According to Definition~\ref{def:lasso-prob}, we know that each lasso occurs with probability $\frac{1}{2}$ and they are not witnesses since the corresponding ultimately periodic words $a^{\omega}$ and $bb^{\omega}$ do not belong to the language $\lang{\baut[A]} \setminus \lang{\baut[B]}$.
	If we set $K = 2$ to check whether $\lang{\baut[A]} \subseteq \lang{\baut[B]}$, we end up concluding that the inclusion holds with probability $1$ since the probability to find some lasso of $\baut[A]$ related to the $\omega$-word $ab^{\omega} \in \lang{\baut[A]} \setminus \lang{\baut[B]}$ is $0$.
	If we want to find a witness $K$-lasso, we need to set $K = 3$ at least, since now the terminating $3$-lasso $\terminatingLasso[s_{1} a s_{1} b s_{2} b s_{2}]$ with corresponding $\omega$-word $abb^{\omega} \in \lang{\baut[A]} \setminus \lang{\baut[B]}$ can be found with probability $\frac{1}{16} > 0$.

	We remark that the Monte Carlo method proposed in~\cite{DBLP:conf/tacas/GrosuS05} uses lassos that are a special instance of Definition~\ref{def:lasso-prob} when we let $K = 2$ and $\stopprob = 1$, thus their method is not complete for NBA language inclusion checking.
\end{myexample}

According to Theorem~\ref{thm:discrete-distribution-inclusion}, the probability space of the sample terminating $K$-lassos in $\baut[A]$ can be organized in the tree, like the one shown in Fig.~\ref{fig:lasso-prob-space-tree-example}.
Therefore, it is easy to see that the probability to find the witness $3$-lasso $\terminatingLasso[s_{1} a s_{1} b s_{2} b s_{2}]$ of $\baut[A]$ is $\frac{1}{16}$, as indicated by the leaf node $\treenode{\terminatingLasso[s_{1} a s_{1} b s_{2} b s_{2}], \frac{1}{16}}$.

\begin{definition}[Lasso Bernoulli Variable]
\label{def:bernoulli-var}
	Let $K\geq 2$ be a natural number and $\stopprob$ a stopping probability.
	The random variable associated with the probability space $(S^{K}_{\baut[A]}, 2^{S^{K}_{\baut[A]}}, \prob_{\stopprob})$ of the NBAs $\baut[A]$ and $\baut[B]$ is defined as follows:
	$p_{Z} = \prob_{\stopprob}[Z = 1] = \sum_{\terminatingLasso \in S_{w}} \prob_{\stopprob}[\terminatingLasso]$ and $q_{Z} = \prob_{\stopprob}[Z = 0] = \sum_{\terminatingLasso \in S_{n}} \prob_{\stopprob}[\terminatingLasso]$, where $S_{w}, S_{n} \subseteq S^{K}_{\baut[A]}$ are the set of witness and non-witness lassos, respectively.
\end{definition}

Under the assumption $\lang{A} \setminus \lang{B} \neq \emptyset$, there exists some witness $K$-lasso $\terminatingLasso \in S_{w}$ that can be sampled with positive probability if $\prob_{\stopprob}[Z = 1] > 0$, as explained by Example~\ref{ex:K-not-large-enough}.

\begin{myexample}
\label{ex:K-not-large-enough}
	For the NBAs $\baut[A]$ and $\baut[B]$ shown in Fig.~\ref{fig:buechi-example}, $K = 3$, and $\stopprob = \frac{1}{2}$, the lasso probability space is organized as in Fig.~\ref{fig:lasso-prob-space-tree-example}.
	The lasso Bernoulli variable has associated probabilities $p_{Z} = \frac{1}{8}$ and $q_{Z} = \frac{7}{8}$ since the only witness lassos are $\terminatingLasso[s_{1} a s_{1} b s_{2} b s_{2}]$ and $\terminatingLasso[s_{1} a s_{1} b s_{2} b s_{2} b s_{2}]$, both occurring with probability $\frac{1}{16}$.
\end{myexample}

Therefore, if we set $K=3$ and $\stopprob = \frac{1}{2}$ to check the inclusion between $\baut[A]$ and $\baut[B]$ from Fig.~\ref{fig:buechi-example}, we are able to find with probability $\frac{1}{8}$ the $\omega$-word $ab^{\omega}$ as a counterexample to the inclusion $\lang{\baut[A]} \subseteq \lang{\baut[B]}$.
It follows that the probability we do not find any witness $3$-lasso after $50$ trials would be less than $0.002$, which can be made even smaller with a larger number of trials.

\begin{figure}[t]
	\centering
	\begin{tikzpicture}[->, >=stealth', auto, initial text={}]
		\path[use as bounding box] (-1,0.45) rectangle (8.5,-1.95);

		\node[initial, state] (q1) {$q_{1}$};
		\node[state] (q2) at ($(q1) + (2,0)$) {$q_{2}$};
		\node[state] (q3) at ($(q2) + (2,0)$) {$q_{3}$};
		\node (qdots) at ($(q3) + (2,0)$) {$\dotso$};
		\node[state] (qK) at ($(qdots) + (2,0)$) {$q_{K}$};
		\node[state, accepting] (qb) at ($0.5*(q1) + 0.5*(qK) + (0,-1.5)$) {$q_{b}$};
		
		\draw (q1) to node[below] {$b$} (qb);
		\draw (q1) to node {$a$} (q2);
		\draw (q2) to node {$b$} (qb);
		\draw (q2) to node {$a$} (q3);
		\draw (q3) to node {$b$} (qb);
		\draw (q3) to node {$a$} (qdots);
		\draw (qdots) to node {$a$} (qK);
		\draw (qK) to node {$b$} (qb);
		
		\draw[loop right] (qb) to node {$b$} (qb);
		
	\end{tikzpicture}
	\caption{NBA $\baut[K]_{K}$ making $p_{Z} = 0$ when checking $\lang{\baut[A]} \subseteq \lang{\baut[K]_{K}}$ by means of sampling terminating $K$-lassos from $\baut[A]$ shown in Fig.~\ref{fig:buechi-example}.}
	\label{fig:b-counterexample}
\end{figure}
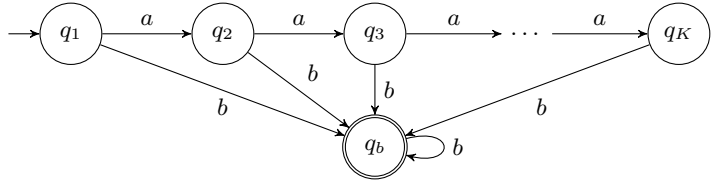

As we have seen in Example~\ref{ex:prob-lassos}, the counterexample may not be sampled with positive probability if $K$ is not sufficiently large, that is the main problem with $\mc$ algorithm from~\cite{DBLP:conf/tacas/GrosuS05} for checking language inclusion.
The natural question is then: how large should $K$ be for checking the inclusion?
First, let us discuss about $K$ without taking the automaton $\baut[B]$ into account.
Consider the NBA $\baut[A]$ of Fig.~\ref{fig:buechi-example}:
it seems that no matter how large $K$ is, one can always construct an NBA $\baut[K]$ with $K + 1$ states to make the probability $p_{Z} = 0$, as the counterexample $a^{l} b^{\omega} \in \lang{\baut[A]} \setminus \lang{\baut[B]}$ can not be sampled for any $l \geq K$.
Fig.~\ref{fig:b-counterexample} depicts such NBA $\baut[K]$, for which we have $\lang{K} = \setnocond{ b^{\omega}, ab^{\omega}, aab^{\omega}, \dotsc, a^{K-1}b^{\omega}}$.
One can easily verify that the counterexample $a^{l}b^{\omega}$ can not be sampled from $\baut[A]$ when $l \geq K$, as sampling this word requires the state $s_{1}$ to occur $l+1$ times in the run, that is not a valid $K$-lasso.
This means that $K$ is a value that depends on the size of $\baut[B]$.
To get a $K$ sufficiently large for every $\baut[A]$ and $\baut[B]$, one can just take the product of $\baut[A]$ with the complement of $\baut[B]$ and check how many times in the worst case a state of $\baut[A]$ occurs in the shortest accepting run of the product.

\begin{restatable}[Sufficiently Large $K$]{lemma}{sufficientlylargeK}
\label{lem:large-K}
	Let $\baut[A]$, $\baut[B]$ be NBAs with $n_{a}$ and $n_{b}$ states, respectively, and $Z$ be the random variable defined in Definition~\ref{def:bernoulli-var}.
	Assume that $\lang{\baut[A]} \setminus \lang{\baut[B]} \neq \emptyset$.
	If $K \geq 2 \times (2n_{b} + 2)^{n_{b}} \times 2^{n_{b}} + 1$, then $\prob_{\stopprob}[Z = 1] > 0$.		
\end{restatable}
\begin{myproof}
To check whether $\lang{\baut[A]} \setminus \lang{\baut[B]} \neq \emptyset$, we can use NBA complementation and product operations to check equivalently $\lang{\baut[A]} \cap \lang{\complement{\baut[B]}} = \emptyset$.
By Lemma~\ref{lem:buechiOperations}, the resulting NBA $\baut[C]$ such that $\lang{\baut[C]} = \lang{\baut[A]} \cap \lang{\complement{\baut[B]}}$ has $2 \times n_{a} \times (2n_{b} + 2)^{n_{b}} \times 2^{n_{b}}$ states, each one of the form $(q_{a}, q_{b}, i)$ with $q_{a} \in \astates_{A}$, $q_{b} \in \astates_{\complement{\baut[B]}}$, and $i \in \setnocond{1, 2}$.
In the worst case, the shortest run $\afrun_{c}$ witnessing $\lang{\baut[C]} \neq \emptyset$ has $\size{\astates_{C}} + 1$ states, where each state of $\baut[C]$ occurs exactly once in $\afrun_{c}$ except for the last one occurring twice.
Since $\baut[C]$ is the product NBA of $\baut[A]$ and $\complement{\baut[B]}$, the run $\afrun_{c}$ can be projected into the component runs $\afrun_{a}$ and $\afrun_{b}$ of $\baut[A]$ and $\complement{\baut[B]}$, respectively (see, e.g.,~\cite{DBLP:reference/mc/Kupferman18});
both of them have the same length as $\afrun_{c}$.
Note that by the product construction, we have that $\afrun_{a}$ and $\afrun_{b}$ are both accepting runs in $\baut[A]$ and $\complement{\baut[B]}$, respectively.
By product and projection construction, we have that each state of $\baut[A]$ occurs exactly $2 \times (2n_{b} + 2)^{n_{b}} \times 2^{n_{b}}$ in $\afrun_{a}$ except for the last one occurring one time more.
This means that by setting $K \geq 2 \times (2n_{b} + 2)^{n_{b}} \times 2^{n_{b}} + 1$, $\afrun_{a}$ becomes a valid $K$-lasso that is sampled with non-zero probability.
\end{myproof}

\begin{remark}
\label{rem:largeKFacts}
	We want to stress that choosing $K$ as given in Lemma~\ref{lem:large-K} is a sufficient condition for sampling a counterexample with positive probability;
	choosing this value, however, is not a necessary condition.
	In practice, we can find counterexamples with positive probability with $K$ being set to a value much smaller than $2 \times (2n_{b} + 2)^{n_{b}} \times 2^{n_{b}} + 1$, as experiments reported in Section~\ref{sec:experiments} indicate.
\end{remark}

\begin{algorithm}[t]
    \caption{$\imc$ Algorithm}
    \label{alg:imc-algo}
    \begin{algorithmic}[1]
		\Procedure{$\imc$}{$\baut[A], \baut[B], K, \stopprob, \errorprob, \confprob$}
		\State {$M := \lceil\ln \confprob / \ln (1 - \errorprob) \rceil$;} \label{alg:imc-algo:line:M}
		\For{($i := 1$; $i \leq M$; $i\mathord{++}$)}
			\State $(u,v) := \mathsf{sample}(\baut[A], K, \stopprob)$; \label{alg:imc-algo:line:sample}
			\If{$\mathsf{membership}(\baut[A], (u,v))$} \label{alg:imc-algo:line:checkA}
				\If{not $\mathsf{membership}(\baut[B], (u,v))$} \label{alg:imc-algo:line:checkB}
					\State{\Return $(\mathit{false}, (u,v))$;} \label{alg:imc-algo:line:returnFalse}
				\EndIf
			\EndIf
		\EndFor
		\State{\Return $\mathit{true}$;} \label{alg:imc-algo:line:returnTrue}
   \EndProcedure
    \end{algorithmic}
\end{algorithm}

Now we are ready to present our $\imc$ algorithm, given in Algorithm~\ref{alg:imc-algo}.
On input the two NBAs $\baut[A]$ and $\baut[B]$, the bound $K$, the stopping probability $\stopprob$, and the statistical parameters $\errorprob$ and $\confprob$, the algorithm at line~\ref{alg:imc-algo:line:M} first computes the number $M$ of samples according to $\errorprob$ and $\confprob$.
Then, for each $\omega$-word $(u,v) = uv^{\omega}$ associated with a terminating lasso sampled at line~\ref{alg:imc-algo:line:sample} according to Definitions~\ref{def:lasso} and~\ref{def:lasso-prob}, it checks whether the lasso is a witness by first (line~\ref{alg:imc-algo:line:checkA}) verifying whether $uv^{\omega} \in \lang{\baut[A]}$, and then (line~\ref{alg:imc-algo:line:checkB}) whether $uv^{\omega} \notin \lang{\baut[B]}$.
If the sampled lasso is indeed a witness, a counterexample to $\lang{\baut[A]} \subseteq \lang{\baut[B]}$ has been found, so the algorithm can terminate at line~\ref{alg:imc-algo:line:returnFalse} with the correct answer $\mathit{false}$ and the counterexample $(u,v)$.
If none of the $M$ sampled lassos is a witness, then the algorithm returns $\mathit{true}$ at line~\ref{alg:imc-algo:line:returnTrue}, which indicates that hypothesis $\lang{\baut[A]} \not\subseteq \lang{\baut[B]}$ has been rejected and $\lang{\baut[A]} \subseteq \lang{\baut[B]}$ is assumed to hold.
It follows that $\imc$ gives a probabilistic guarantee in terms of finding a counterexample to inclusion when $K$ is sufficient large, as formalized by the following proposition.
\begin{proposition}
	Let $\baut[A]$, $\baut[B]$ be two NBAs and $K$ be a sufficiently large number.
	If $\lang{\baut[A]} \setminus \lang{\baut[B]} \neq \emptyset$, then $\imc$ finds a counterexample to the inclusion $\lang{\baut[A]} \subseteq \lang{\baut[B]}$ with positive probability.
\end{proposition}

In general, the exact value of $p_{Z}$, the probability of finding a word accepted by $\baut[A]$ but not accepted by $\baut[B]$, is unknown or at least very hard to compute.
Thus, we summarize our results about $\imc$ in Theorems~\ref{thm:imc-correctness} and~\ref{thm:imc-complexity} with respect to the choice of the statistical parameters $\errorprob$ and $\confprob$.
\begin{restatable}[Correctness]{theorem}{imcCorrectness}
\label{thm:imc-correctness}
	Let $\baut[A]$, $\baut[B]$ be two NBAs, $K$ be a sufficiently large number, and $\errorprob$ and $\confprob$ be statistical parameters.
	If $\imc$ returns $\mathit{false}$, then $\lang{\baut[A]} \not\subseteq \lang{\baut[B]}$ is certain.
	Otherwise, if $\imc$ returns $\mathit{true}$, then the probability that we would continue and with probability $p_{Z} \geq \errorprob$ find a counterexample is less than $\confprob$.
\end{restatable}

\begin{myproof}
	The fact that $\imc$ is correct when returning $\mathit{false}$ is trivial, since this happens only when $\imc$ finds an $\omega$-word $uv^{\omega}$ such that $uv^{\omega} \in \lang{\baut[A]}$ (cf. line~\ref{alg:imc-algo:line:checkA}) and $uv^{\omega} \notin \lang{\baut[B]}$ (cf. line~\ref{alg:imc-algo:line:checkB}), i.e., $uv^{\omega}$ is a witness of $\lang{\baut[A]} \not\subseteq \lang{\baut[B]}$.
	
	The fact that $\imc$, on returning $\mathit{true}$, ensures that, by sampling more words, with probability at most $\confprob$ a counterexample is found with probability $p_{Z} \geq \errorprob$, is justified by statistical hypothesis testing (cf.~Section~\ref{ssec:pre-sampling-testing}).
\end{myproof}

\begin{restatable}[Complexity]{theorem}{imcComplexity}
\label{thm:imc-complexity}
	Given two NBAs $\baut[A]$, $\baut[B]$ with $n_{a}$ and $n_{b}$ states, respectively, and statistical parameters $\errorprob$ and $\confprob$, let $M = \lceil\ln \confprob / \ln (1 - \errorprob) \rceil$ and $n = \max(n_{a}, n_{b})$.
	Then $\imc$ runs in time $\bigO(M \cdot K \cdot n^{2})$ and space $\bigO(K \cdot n^{2})$.
\end{restatable}
\begin{myproof}
	Let $(u, v)$ be a sampled word from $\baut[A]$.
	Then in the worst case, $\size{u} + \size{v} = n \times (K - 1)$ when every state in $\baut[A]$ occurs in the sampled lasso $K - 1$ times, with one state occurring $K$ times.
	According to Lemma~\ref{lem:nba-membership}, determining whether an $\omega$-word $(u, v) = uv^{\omega}$ is accepted by an NBA can be done in time and space linear in the number of states and the length of $(u, v)$.
	Therefore the time and space complexity for resolving a membership checking problem of $(u, v)$ is in $\bigO(n \times K \times n) = \bigO(K \cdot n^{2})$.
	It follows that $\imc$ runs in time $\bigO(M \cdot K \cdot n^{2})$ and space $\bigO(K \cdot n^{2})$.
\end{myproof}

\section{Experimental Evaluation}
\label{sec:experiments}

We have implemented the Monte Carlo sampling algorithm proposed in Section~\ref{sec:algorithm} in \roll~\cite{DBLP:conf/tacas/LiSTCX19} to evaluate it.
We performed our experiments on a desktop PC equipped with a $3.6$ GHz Intel i7-4790 processor with $16$ GB of RAM, of which $4$ GB were assigned to the tool.
We imposed a timeout of $300$ seconds ($5$ minutes) for each inclusion test.
In the experiments, we compare our sampling inclusion test algorithm with \rabit  2.4.5~\cite{DBLP:conf/cav/AbdullaCCHHMV10,DBLP:conf/concur/AbdullaCCHHMV11,DBLP:journals/lmcs/ClementeM19} and \spot 2.8.4~\cite{DBLP:conf/atva/Duret-LutzLFMRX16}.
\roll and \rabit are written in Java while \spot is written in C/C++.
This gives \spot some advantage in the running time, since it avoids the overhead caused by the Java Virtual Machine.
For \rabit we used the option \texttt{-fastc} while for \roll we set parameters $\errorprob = 0.1\%$ and $\confprob = 2\%$, resulting in sampling roughly $4\,000$ words for testing inclusion, $\stopprob = \frac{1}{2}$, and $K$ to the maximum of the number of states of the two automata.
The automata we used in the experiment are represented in two formats:
the BA format used by \goal\footnote{\goal is omitted in our experiments as it is shown in~\cite{DBLP:journals/lmcs/ClementeM19} that \rabit performs much better than \goal.}~\cite{DBLP:conf/cav/TsaiTH13} and the HOA format~\cite{DBLP:conf/cav/BabiakBDKKM0S15}.
\rabit supports only the former, \spot only the latter, while \roll supports both.
We used \roll to translate between the two formats and then we compared \roll (denoted \rollH) with \spot on the HOA format and \roll (denoted \rollB) with \rabit on the BA format.
When we present the outcome of the experiments, we distinguish them depending on the used automata format.
This allows us to take into account the possible effects of the automata representation, on both the language they represent and the running time of the tools.

\subsection{Experiments on Randomly Generated \buechi Automata}
\label{ssec:experimentsRandom}

\begin{table}[t]
	\caption{Experiment results on random automata with fixed state space and alphabet.}
	\label{tab:experimentsRandomOverviewStatesAlphabet}
	\centering
	\begin{tabular}{l|P{3}+{5}rrr}
		\multicolumn{1}{c|}{Tool} & \multicolumn{1}{c}{included} & \multicolumn{1}{c}{not included} & timeout & memory out & other failures \\
		\hline
		\spot & 1\,803 & 10\,177+53 & 1\,780 & 670 & 1\,517\\
		\rollH & 2\,497(5) & 10\,177+3\,194 & 119 & 0 & 13\\
		\hhline
		\rollB & 2\,501(45) & 12\,436+1\,054 & 0 & 0 & 9\\
		\rabit & 2\,205 & 12\,436+45 & 306 & 1\,008 & 0
	\end{tabular}
\end{table}
To run the different tools on randomly generated automata, we used \spot to generate $50$ random HOA automata for each combination of state space size $\size{\astates} \in \setnocond{10, 20, \dotsc, 90, 100, 125, \dotsc, 225, 250}$ and alphabet size $\size{\alphabet} \in \setnocond{2, 4, \dotsc, 18, 20}$, for a total of $8\,000$ automata, that we have then translated to the BA format.
We then considered $100$ different pairs of automata for each combination of state space size and alphabet size (say, for instance, $100$ pairs of automata with $50$ states and $10$ letters or $100$ pairs with $175$ states and $4$ letters).
The resulting $16\,000$ experiments are summarized in Table~\ref{tab:experimentsRandomOverviewStatesAlphabet}.

For each tool, we report the number of inclusion test instances that resulted in an answer for language inclusion and not inclusion, as well as the number of cases where a tool went timeout, ran out of memory, or failed for any other reason.
For the ``included'' case, we indicate in parenthesis how many times \roll has failed to reject the hypothesis $\lang{\baut[A]} \subseteq \lang{\baut[B]}$, that is, \roll returned ``included'' instead of the expected ``not included''.
For the ``non included'' case, instead, we split the number of experiments on which multiple tools returned ``not included'' and the number of times only this tool returned ``not included'';
for instance, we have that both \spot and \rollH returned ``not included'' on $10\,177$ cases, that only \spot returned so in $53$ more experiments (for a total of $10\,230$ ``not included'' results), and that only \rollH identified non inclusion in $3\,194$ additional experiments (for a total of $13\,371$ ``not included'' results).

We can see in Table~\ref{tab:experimentsRandomOverviewStatesAlphabet} that both \rollH and \rollB were able to solve many more cases than their counterparts \spot and \rabit, respectively, on both ``included'' and ``not included'' outcomes.
In particular, we can see that both \rollH and \rollB have been able to find a counterexample to the inclusion for many cases ($3\,194$ and $1\,052$, respectively) where \spot on the HOA format and \rabit on the BA format failed, respectively.

On the other hand, there are only few cases where \spot or \rabit proved non inclusion while \roll failed to do so.
In particular, since \roll implements a statistical hypothesis testing algorithm for deciding language inclusion, we can expect few experiments where \roll fails to reject the alternative hypothesis $\lang{\baut[A]} \subseteq \lang{\baut[B]}$.
In the experiments this happened $5$ (\rollH) and $45$ (\rollB) times;
this corresponds to a failure rate of less than $0.6\%$, well below the choice of the statistical parameter $\confprob = 2\%$.

Regarding the $13$ failures of \rollH and the $9$ ones of \rollB, they are all caused by a stack overflow in the strongly connected components (SCC) decomposition procedure for checking membership $uv^{\omega} \in \lang{\baut[A]}$ or $uv^{\omega} \in \lang{\baut[B]}$ (i.e., $\lang{\baut[A]} \cap \{uv^{\omega}\} = \emptyset$ or $\lang{\baut[B]} \cap \{uv^{\omega}\} = \emptyset$, cf.~\cite{DBLP:reference/mc/Kupferman18}) at lines~\ref{alg:imc-algo:line:checkA} and~\ref{alg:imc-algo:line:checkB} of Algorithm~\ref{alg:imc-algo}, since checking whether the sampled lasso is an accepting run of $\baut[A]$ does not suffice (cf. Remark~\ref{rem:KlassoProperties}).
The $119$ timeouts of \rollH occurred for $3$ pairs of automata with $200$ states and $20$ letters, $12$/$21$ pairs of automata with $225$ states and $18$/$20$ letters, respectively, and $40$/$43$ pairs of automata with $250$ states and $18$/$20$ letters, respectively.
We plan to investigate why \rollH suffered of these timeouts while \rollB avoided them, to improve \roll's performance.

\begin{figure}[t]
	\centering
	\includegraphics{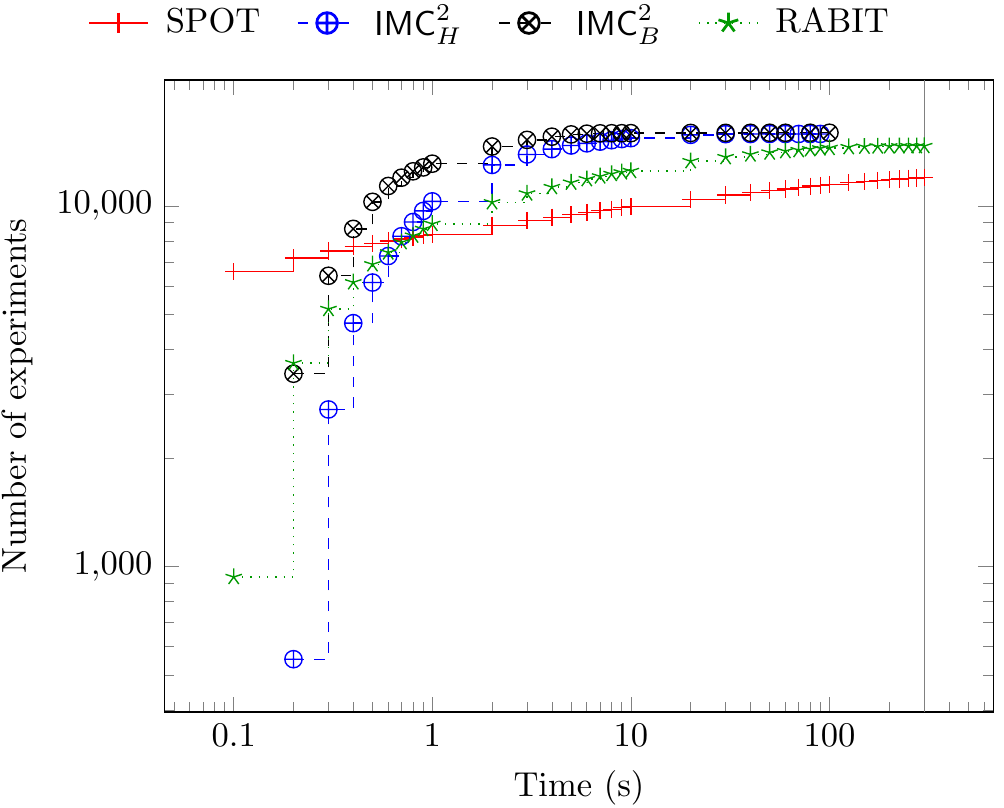}
	\caption{Experiment running time on the random automata with fixed state space and alphabet.}
	\label{fig:experimentsRandomOverviewStatesRuntime}
\end{figure}

About the execution running time of the tools, they are usually rather fast in giving an answer, as we can see from the plot in Fig.~\ref{fig:experimentsRandomOverviewStatesRuntime}.
In this plot, we show on the $y$ axis the total number of experiments, each one completed within the time marked on the $x$ axis;
the vertical gray line marks the timeout limit.
The plot is relative to the number of ``included'' and ``not included'' outcomes combined together;
the shape of the plots for the two outcomes kept separated is similar to the combined one we present in Fig.~\ref{fig:experimentsRandomOverviewStatesRuntime};
the only difference is that in the ``not included'' case, the plots for \rollB and \rollH would terminate earlier, since all experiments returning ``not included'' are completed within a smaller time than for the ``included'' case.
As we can see, we have that \roll rather quickly overcame the other tools in giving an answer.
This is likely motivated by the fact that by using randomly generated automata, the structure-based tools such as \rabit and \spot are not able to take advantage of the symmetries or other structural properties one can find in automata obtained from, e.g., logical formulas.
From the result of the experiments presented in Table~\ref{tab:experimentsRandomOverviewStatesAlphabet} and Fig.~\ref{fig:experimentsRandomOverviewStatesRuntime}, we have that the use of a sampling-based algorithm is a very fast, effective, and reliable way to rule out that $\lang{\baut[A]} \subseteq \lang{\baut[B]}$ holds.
Moreover, we also conclude that $\imc$ complements existing approaches rather well, as it finds counterexamples to the language inclusion for a lot of instances that other approaches fail to manage.

\subsection{Effect of the Statistical Parameters $\errorprob$ and $\confprob$}
\label{ssec:experimentsParameters}

To analyze the effect of the choice of $\errorprob$ and $\confprob$ on the execution of the sampling algorithm we proposed, we have randomly taken $100$ pairs of automata where, for each pair $(\baut[A], \baut[B])$, the automata $\baut[A]$ and $\baut[B]$ have the same alphabet but possibly different state space.
On these $100$ pairs of automata, we repeatedly ran \rollH $10$ times with different values of $\errorprob$ in the set $\setnocond{0.00001, 0.00051, \dotsc, 0.00501}$ and of $\confprob$ in the set $\setnocond{0.0001, 0.0051, \dotsc, 0.0501}$, for a total of $121\,000$ inclusion tests.

The choice of $\errorprob$ and $\confprob$ plays essentially no role in the running time for the cases where a counterexample to the language inclusion is found: the average running time is between $1.67$ and $1.77$ seconds.
This can be expected, since \roll stops its sampling procedure as soon as a counterexample is found (cf. Algorithm~\ref{alg:imc-algo}).
If we consider the number of experiments, again there is almost no difference, since for all combinations of the parameters it ranges between $868$ and $870$.

On the other hand, $\errorprob$ and $\confprob$ indeed affect the running time for the ``included'' cases, since they determine the number $M$ of sampled words and all such words have to be sampled and tested before rejecting the ``non included'' hypothesis.
The average running time is $1$ second or less for all choices of $\errorprob \neq 0.00001$ and $\confprob$, while for $\errorprob = 0.00001$, the average running time ranges between $12$ and $36$ seconds when $\confprob$ moves from $0.0501$ to $0.0001$, which corresponds to testing roughly $300\,000$ to $1\,000\,000$ sample words, respectively.

\subsection{Effect of the Lasso Parameters $K$ and $\stopprob$}
\label{ssec:experimentsK}

At last, we also experimented with different values of $K$ and $\stopprob$ while keeping the statistical parameters unchanged:
we have generated other $100$ pairs of automata as in Section~\ref{ssec:experimentsParameters} and then checked inclusion $10$ times for each pair and each combination of $K \in \setnocond{2, 3, 4, 5, 6, 8, 11, 51, 101, 301}$ and $\stopprob \in \setnocond{0.05, 0.1, \dotsc, 0.95}$.

As one can expect, low values for $\stopprob$ and large values of $K$ allow $\imc$ to find more counterexamples, at the cost of a higher running time.
It is worth noting that $K = 2$ is still rather effective in finding counterexamples: 
out of the $1\,000$ executions on the pairs, $\imc$ returned ``non included'' between $906$ and $910$ times;
for $K = 3$ it ranged between $914$ and $919$ for $\stopprob \leq 0.5$ and between $909$ and $912$ for $\stopprob > 0.5$.
Larger values of $K$ showed similar behavior.
Regarding the running time, except for $K=2$ the running time of $\imc$ is loosely dependent on the choice of $K$, for a given $\stopprob$;
this is likely motivated by the fact that imposing e.g. $K=51$ still allows $\imc$ to sample lassos that are for instance $4$-lassos.
Instead, the running time is affected by the choice of $\stopprob$ for a given $K \geq 3$: 
as one can expect, the smaller $\stopprob$ is, the longer $\imc$ takes to give an answer;
a small $\stopprob$ makes the sampled words $uv^{\omega} \in \lang{\baut_{1}}$ to be longer, which in turn makes the check $uv^{\omega} \in \lang{\baut_{2}}$ more expensive.

Experiments suggest that taking $0.25 \leq \stopprob \leq 0.5$ and $3 \leq K \leq 11$ gives a good tradeoff between running time and number of ``non included'' outcomes.
Very large values of $K$, such as $K > 50$, are usually not needed, also given the fact that usually lassos with several repetitions occur with rather low probability.

\section{Conclusion and Discussion}
\label{sec:conclusion}

We presented $\imc$, a sample-based algorithm  for proving language non-inclusion between \buechi automata.
Experimental evaluation showed that $\imc$ is very fast and reliable in finding such witnesses, by sampling them in many cases where traditional structure-based algorithms fail or take too long to complete the analysis.
We believe that $\imc$ is a very good technique to disprove $\lang{\baut[A]} \subseteq \lang{\baut[B]}$ and complements well the existing techniques for checking \buechi automata language inclusion.
As future work, our algorithm can be applied to scenarios like black-box testing and PAC learning~\cite{DBLP:journals/ml/Angluin87}, in which inclusion provers are either not applicable in practice or not strictly needed.
A uniform word sampling algorithm was proposed in~\cite{DBLP:conf/concur/BassetMS17} for concurrent systems with multiple components. 
We believe that extending our sampling algorithms to concurrent systems with multiple components is worthy of study.

\bibliographystyle{splncs04}
\bibliography{bibreduced}

\appendix

\section{Proof of Theorem~\ref{thm:discrete-distribution-inclusion}}
\label{app:proofs}

\lassoprobspace*

The fact that $\prob(\afrun)$ is a non-negative real value for each $\afrun \in S$ is immediate by definition of $\prob$.
By definition, $\prob$ satisfies $\prob_{\stopprob}[S_{1} \cup S_{2}] = \prob_{\stopprob}[S_{1}] + \prob_{\stopprob}[S_{2}]$ for each $S_{1}, S_{2} \subseteq S^{K}_{\baut[A]}$ such that $S_{1} \cap S_{2} = \emptyset$.
To complete the proof, we need to show that $\prob_{\stopprob}[S^{K}_{\baut[A]}] = 1$.

Let $P = \setcond{\afrun' \in \astates \times (\alphabet \times \astates)^{*}}{\text{$\afrun'$ is a prefix of some $\terminatingLasso \in S^{K}_{\baut[A]}$}}$ be the set of proper prefixes of the $K$-lassos in $S^{K}_{\baut[A]}$.
For a given finite run $\afrun = s_{0} a_{0} s_{1} \dotsc a_{n-1} s_{n}$, let $\afrun_{l}$ denote the last state $s_{n}$ of $\afrun$.
Consider the tree $\tree = (N, \treenode{\emptyword, 1}, E)$ where
\begin{itemize}
\item
	$N = \big(P \times (0, 1]\big) \cup \big(S^{K}_{\baut[A]} \times (0, 1]\big) \cup \setnocond{\treenode{\emptyword, 1}}$ is the set of nodes,
\item
	$\treenode{\emptyword, 1}$ is the root of the tree, and
\item
	$E \subseteq \Big(\setnocond{\treenode{\emptyword, 1}} \times \big(P \times (0, 1]\big)\Big) \cup \Big(P \times (0, 1]\Big)^{2} \cup \Big(\big(P \times (0, 1]\big) \times \big(S^{K}_{\baut[A]} \times (0, 1]\big)\Big)$ is the set of edges defined as
	\begin{align*}
		E = & \hphantom{{} \cup {}}\setcond{(\treenode{\emptyword, 1}, \treenode{q, \frac{1}{\size{\ainit}}})}{q \in \ainit} \\
		& {} \cup \setcond{\big(\treenode{\afrun, p}, \treenode{\afrun a q, \frac{p}{\size{\amat(\afrun_{l})}}}\big)}{\afrun a q \in P \land \appear{\afrun}{\afrun_{l}} = 1} \\
		& {} \cup \setcond{\big(\treenode{\afrun, p}, \treenode{\afrun a q,\frac{p \cdot (1 - \stopprob)}{\size{\amat(\afrun_{l})}}}\big)}{\afrun a q \in P \land \appear{\afrun}{\afrun_{l}} > 1} \\
		& {} \cup \setcond{\big(\treenode{\afrun, p}, \treenode{\terminatingLasso, p}\big)}{\terminatingLasso \in S^{K}_{\baut[A]} \land \appear{\afrun}{\afrun_{l}} = K} \\
		& {} \cup \setcond{\big(\treenode{\afrun, p}, \treenode{\terminatingLasso, p \cdot \stopprob}\big)}{\terminatingLasso \in S^{K}_{\baut[A]} \land \appear{\afrun}{\afrun_{l}} < K}
	\end{align*}
\end{itemize}
Note that by construction, $\tree$ is finite.

We now show by induction on the length of the finite run $\afrun \in P$ that $\treenode{\afrun, \prob'_{\stopprob}[\afrun]} \in N$ is reachable in $\tree$ from $\treenode{\emptyword, 1}$.
\begin{description}
\item[Base case $\afrun = q_{0}$:]
	by definition of $E$, we have that $(\treenode{\emptyword, 1}, \treenode{q_{0}, \frac{1}{\size{\ainit}}}) \in E$ since by Definition~\ref{def:lasso} we have that $q_{0} \in \ainit$, which implies that $(\treenode{\emptyword, 1}, \treenode{q_{0}, \prob'_{\stopprob}[q_{0}]}) \in E$ since by Definition~\ref{def:lasso-prob} we have that $\prob'[q_{0}] = \frac{1}{\size{\ainit}}$, thus  $(\treenode{\emptyword, 1}, \treenode{q_{0}, \prob'_{\stopprob}[\afrun]}) \in E$ showing that $\treenode{\afrun, \prob'_{\stopprob}[\afrun]} \in N$ is reachable in $\tree$ from $\treenode{\emptyword, 1}$.
\item[Inductive step:]
	let $\afrun$ be $\afrun' a q$ for some $\afrun' \in P$.
	By induction hypothesis, we have that $\treenode{\afrun', \prob'_{\stopprob}[\afrun']}$ is reachable in $\tree$ from $\treenode{\emptyword, 1}$.
	There are now two cases:
	either $\appear{\afrun'}{\afrun'_{l}} = 1$ or $\appear{\afrun'}{\afrun'_{l}} > 1$.
	
	If $\appear{\afrun'}{\afrun'_{l}} = 1$, then by definition of $E$ we have that $\big(\treenode{\afrun', \prob'_{\stopprob}[\afrun']}, \treenode{\afrun' a q, \frac{\prob'_{\stopprob}[\afrun']}{\size{\amat{\afrun'_{l}}}}}\big) \in E$, which by simple rewriting is $\big(\treenode{\afrun', \prob'_{\stopprob}[\afrun']}, \treenode{\afrun' a q, \prob'_{\stopprob}[\afrun'] \cdot \frac{1}{\size{\amat{\afrun'_{l}}}}}\big) \in E$ that can be rewritten as $\big(\treenode{\afrun', \prob'_{\stopprob}[\afrun']}, \treenode{\afrun' a q, \prob'_{\stopprob}[\afrun'] \cdot \pi[\afrun'_{l} a q]}\big) \in E$ since $(\afrun'_{l}, a, q) \in \amat$ derives from the fact that $\afrun' a q$ is a prefix of some $K$-lasso in $S^{K}_{\baut[A]}$ and $\pi[\afrun'_{l} a q] = \frac{1}{m}$ with $m = \size{\amat(\afrun'_{l})}$ from Definition~\ref{def:lasso-prob}, which also implies that $\big(\treenode{\afrun', \prob'_{\stopprob}[\afrun']}, \treenode{\afrun, \prob'_{\stopprob}[\afrun]}\big) \in E$ as required, since $\afrun = \afrun' a q$ and $\prob'_{\stopprob}[\afrun] = \prob'_{\stopprob}[\afrun'] \cdot \pi[\afrun'_{l} a q]$, that is, $\treenode{\afrun, \prob'_{\stopprob}[\afrun]} \in N$ is reachable in $\tree$ from $\treenode{\emptyword, 1}$.
	
	Suppose now that $\appear{\afrun'}{\afrun'_{l}} > 1$, then by definition of $E$ we have that $\big(\treenode{\afrun', \prob'_{\stopprob}[\afrun']}, \treenode{\afrun' a q, \frac{\prob'_{\stopprob}[\afrun'] \cdot (1 - \stopprob)}{\size{\amat{\afrun'_{l}}}}}\big) \in E$, which by simple rewriting is $\big(\treenode{\afrun', \prob'_{\stopprob}[\afrun']}, \treenode{\afrun' a q, (1 - \stopprob) \cdot \prob'_{\stopprob}[\afrun'] \cdot \frac{1}{\size{\amat{\afrun'_{l}}}}}\big) \in E$ that can be rewritten as $\big(\treenode{\afrun', \prob'_{\stopprob}[\afrun']}, \treenode{\afrun' a q, (1 - \stopprob) \cdot \prob'_{\stopprob}[\afrun'] \cdot \pi[\afrun'_{l} a q]}\big) \in E$ since $(\afrun'_{l}, a, q) \in \amat$ derives from the fact that $\afrun' a q$ is a prefix of some $K$-lasso in $S^{K}_{\baut[A]}$ and $\pi[\afrun'_{l} a q] = \frac{1}{m}$ with $m = \size{\amat(\afrun'_{l})}$ from Definition~\ref{def:lasso-prob}, which also implies that $\big(\treenode{\afrun', \prob'_{\stopprob}[\afrun']}, \treenode{\afrun, \prob'_{\stopprob}[\afrun]}\big) \in E$ as required, since $\afrun = \afrun' a q$ and $\prob'_{\stopprob}[\afrun] = (1 - \stopprob) \cdot \prob'[\afrun'] \cdot \pi[\afrun'_{l} a q]$, that is, $\treenode{\afrun, \prob'_{\stopprob}[\afrun]} \in N$ is reachable in $\tree$ from $\treenode{\emptyword, 1}$.
\end{description}
This concludes the proof that for each finite run $\afrun \in P$ we have that $\treenode{\afrun, \prob'_{\stopprob}[\afrun]} \in N$ is reachable in $\tree$ from $\treenode{\emptyword, 1}$.

An analogous result holds also for each terminating $K$-lasso:
for each terminating $K$-lasso $\terminatingLasso \in S^{K}_{\baut[A]}$ we have that $\treenode{\terminatingLasso, \prob_{\stopprob}[\terminatingLasso]} \in N$ is reachable in $\tree$ from $\treenode{\emptyword, 1}$.
Let $\terminatingLasso \in S^{K}_{\baut[A]}$.
By definition of $P$, we have that $\afrun \in P$, so by the result shown above we have that $\treenode{\afrun, \prob'_{\stopprob}[\afrun]} \in N$ is reachable in $\tree$ from $\treenode{\emptyword, 1}$.
To complete the proof, we just need to show that $\big(\treenode{\afrun, \prob'_{\stopprob}[\afrun]}, \treenode{\terminatingLasso, \prob_{\stopprob}[\terminatingLasso]}\big) \in E$.
There are now two cases:
either $\appear{\afrun}{\afrun_{l}} = K$ or $\appear{\afrun}{\afrun_{l}} < K$.
In the former case $\appear{\afrun}{\afrun_{l}} = K$, the definition of $E$ implies that $\big(\treenode{\afrun, \prob'_{\stopprob}[\afrun]}, \treenode{\terminatingLasso, \prob'_{\stopprob}[\afrun]}\big) \in E$ which is indeed $\big(\treenode{\afrun, \prob'_{\stopprob}[\afrun]}, \treenode{\terminatingLasso, \prob_{\stopprob}[\terminatingLasso]}\big) \in E$ since Definition~\ref{def:lasso-prob} implies that $\prob_{\stopprob}[\terminatingLasso] = \prob'_{\stopprob}[\afrun]$, as required.
Similarly, in the latter case $\appear{\afrun}{\afrun_{l}} < K$, the definition of $E$ implies that $\big(\treenode{\afrun, \prob'_{\stopprob}[\afrun]}, \treenode{\terminatingLasso, \stopprob \cdot \prob'_{\stopprob}[\afrun]}\big) \in E$ which is indeed $\big(\treenode{\afrun, \prob'_{\stopprob}[\afrun]}, \treenode{\terminatingLasso, \prob_{\stopprob}[\terminatingLasso]}\big) \in E$ since Definition~\ref{def:lasso-prob} implies that $\prob_{\stopprob}[\terminatingLasso] = \stopprob \cdot \prob'_{\stopprob}[\afrun]$, as required.

By a completely symmetric reasoning, we can show that each node $\treenode{\afrun, p} \in P \times (0, 1]$ that is reachable in $\tree$ from $\treenode{\emptyword, 1}$ has $p = \prob'_{\stopprob}[\afrun]$ and that each node $\treenode{\terminatingLasso, p} \in S^{K}_{\baut[A]} \times (0, 1]$ that is reachable in $\tree$ from $\treenode{\emptyword, 1}$ has $p = \prob_{\stopprob}[\terminatingLasso]$.

These results allow us to claim that checking $\sum_{\terminatingLasso \in S^{K}_{\baut[A]}} \prob_{\stopprob}[\terminatingLasso] = 1$ is equivalent to check $\sum_{\treenode{\terminatingLasso, p}} p = 1$ where the summation is taken over all leaf nodes $\treenode{\terminatingLasso, p} \in N$ that are reachable in $\tree$ from $\treenode{\emptyword, 1}$.

To show this last point, we now prove that for each non-leaf node $\treenode{\afrun, p} \in \big(P \times (0, 1]\big) \cup \setnocond{\treenode{\emptyword, 1}}$, it holds that $\sum_{\treenode{\afrun', p'}: (\treenode{\afrun, p}, \treenode{\afrun', p'}) \in E} p' = p$.
Since the reachable part of the tree $\tree$ is finite, this result allows us to propagate backward, level by level, the probability values stored in the leaves, i.e., on the elements of $S^{K}_{\baut[A]} \times (0,1]$, by adding them together and equating them with the probability value stored in their direct predecessor.
At the end of this propagation, i.e., when we reach the root of $\tree$, we will have that $\sum_{\treenode{\terminatingLasso, p}} p = p_{r}$ with $p_{r}$ being the value associated with the root, that is, $p_{r} = 1$ as required since the root of $\tree$ is $\treenode{\emptyword, 1}$.

Let $\treenode{\afrun, p} \in \big(P \times (0, 1]\big) \cup \setnocond{\treenode{\emptyword, 1}}$ be an arbitrary non-leaf node of $\tree$;
we want to prove that $\sum_{\treenode{\afrun', p'}: (\treenode{\afrun, p}, \treenode{\afrun', p'}) \in E} p' = p$.
There are two cases:
$\treenode{\afrun, p} = \treenode{\emptyword, 1}$ or $\treenode{\afrun, p} \in P \times (0, 1]$.
In the former case, we have that
\begin{align*}
	\sum_{\treenode{\afrun', p'}: (\treenode{\afrun, p}, \treenode{\afrun', p'}) \in E} p'
	& = \sum_{\treenode{q, \frac{1}{\size{\ainit}}}: (\treenode{\emptyword, 1}, \treenode{q, \frac{1}{\size{\ainit}}}) \in E} \frac{1}{\size{\ainit}} \\
	& = \size{\ainit} \cdot \frac{1}{\size{\ainit}} \\
	& = 1\text{,}
\end{align*}
as required, since by definition of $E$, we have an edge $(\treenode{\emptyword, 1}, \treenode{q, \frac{1}{\size{\ainit}}}) \in E$ for each $q \in \ainit$.
In the latter case, i.e., when $\treenode{\afrun, p} \in P \times (0, 1]$, we have that
\begin{align*}
	& \sum_{\treenode{\afrun', p'}: (\treenode{\afrun, p}, \treenode{\afrun', p'}) \in E} p' \\
	& {} = \hphantom{{} + {}} \sum_{\treenode{\afrun a q, \frac{p}{\size{\amat(\afrun_{l})}}}: (\treenode{\afrun, p}, \treenode{\afrun a q, \frac{p}{\size{\amat(\afrun_{l})}}}) \in E \land \afrun a q \in P \land \appear{\afrun}{\afrun_{l}} = 1} \frac{p}{\size{\amat(\afrun_{l})}}\\
	& \hphantom{{} = {}} + \sum_{\treenode{\afrun a q, \frac{p \cdot (1 - \stopprob)}{\size{\amat(\afrun_{l})}}}: (\treenode{\afrun, p}, \treenode{\afrun a q, (1 - \stopprob) \cdot \frac{p}{\size{\amat(\afrun_{l})}}}) \in E \land \afrun a q \in P \land \appear{\afrun}{\afrun_{l}} > 1} (1 - \stopprob) \cdot \frac{p}{\size{\amat(\afrun_{l})}}\\
	& \hphantom{{} = {}} + \sum_{\treenode{\terminatingLasso, p}: (\treenode{\afrun, p}, \treenode{\terminatingLasso, p}) \in E \land \terminatingLasso \in S^{K}_{\baut[A]} \land \appear{\afrun}{\afrun_{l}} = K} p \\
	& \hphantom{{} = {}} + \sum_{\treenode{\terminatingLasso, p \cdot \stopprob}: (\treenode{\afrun, p}, \treenode{\terminatingLasso, p \cdot \stopprob}) \in E \land \terminatingLasso \in S^{K}_{\baut[A]} \land \appear{\afrun}{\afrun_{l}} < K} p \cdot \stopprob\text{.}
\end{align*}
By definition of $E$, the four summations above are on disjoint sets;
moreover, the first and third sets make the other sets to be empty:
if $\setcond{\big(\treenode{\afrun, p}, \treenode{\afrun a q, \frac{p}{\size{\amat(\afrun_{l})}}}\big)}{\afrun a q \in P \land \appear{\afrun}{\afrun_{l}} = 1} \neq \emptyset$ then clearly $\setcond{\big(\treenode{\afrun, p}, \treenode{\afrun a q, \frac{p}{\size{\amat(\afrun_{l})}}}\big)}{\afrun a q \in P \land \appear{\afrun}{\afrun_{l}} > 1} = \emptyset$ and $\setcond{\big(\treenode{\afrun, p}, \treenode{\terminatingLasso, p}\big)}{\terminatingLasso \in S^{K}_{\baut[A]} \land \appear{\afrun}{\afrun_{l}} = K} = \emptyset$ since $\appear{\afrun}{\afrun_{l}} = 1$ contradicts both $\appear{\afrun}{\afrun_{l}} > 1$ and $\appear{\afrun}{\afrun_{l}} = K$ given that $K \geq 2$ by assumption.
Moreover, $\appear{\afrun}{\afrun_{l}} = 1$ forces $\setcond{\big(\treenode{\afrun, p}, \treenode{\terminatingLasso, p}\big)}{\terminatingLasso \in S^{K}_{\baut[A]} \land \appear{\afrun}{\afrun_{l}} < K} = \emptyset$ since Definition~\ref{def:lasso} requires $\appear{\afrun}{\afrun_{l}} \geq 2$ in order to have $\terminatingLasso \in S^{K}_{\baut[A]}$.

This allows us to consider the three cases independently.
\begin{description}
\item[Case $\setcond{\big(\treenode{\afrun, p}, \treenode{\afrun a q, \frac{p}{\size{\amat(\afrun_{l})}}}\big)}{\afrun a q \in P \land \appear{\afrun}{\afrun_{l}} = 1} \neq \emptyset$:]
	we have that
	\begin{align*}
		& \sum_{\treenode{\afrun', p'}: (\treenode{\afrun, p}, \treenode{\afrun', p'}) \in E} p' \\
		& {} = \sum_{\treenode{\afrun a q, \frac{p}{\size{\amat(\afrun_{l})}}}: (\treenode{\afrun, p}, \treenode{\afrun a q, \frac{p}{\size{\amat(\afrun_{l})}}}) \in E \land \afrun a q \in P \land \appear{\afrun}{\afrun_{l}} = 1} \frac{p}{\size{\amat(\afrun_{l})}} \\
		& {} = \size{\amat(\afrun_{l})} \cdot \frac{p}{\size{\amat(\afrun_{l})}} \\
		& {} = p
	\end{align*}
	as required, since by definition of $E$, we have an edge $(\treenode{\afrun, p}, \treenode{\afrun a q, \frac{p}{\size{\amat(\afrun_{l})}}}) \in E$ for each $(\afrun_{l}, a, q) \in \amat(\afrun_{l})$.
\item[Case $\setcond{\big(\treenode{\afrun, p}, \treenode{\terminatingLasso, p}\big)}{\terminatingLasso \in S^{K}_{\baut[A]} \land \appear{\afrun}{\afrun_{l}} = K} \neq \emptyset$:]
	we have that
	\begin{align*}
		& \sum_{\treenode{\afrun', p'}: (\treenode{\afrun, p}, \treenode{\afrun', p'}) \in E} p' \\
		& {} = \sum_{\treenode{\terminatingLasso, p}: (\treenode{\afrun, p}, \treenode{\terminatingLasso, p}) \in E \land \terminatingLasso \in S^{K}_{\baut[A]} \land \appear{\afrun}{\afrun_{l}} = K} p \\
		& {} = p
	\end{align*}
	as required, since by definition of $E$ and $S^{K}_{\baut[A]}$, we have exactly one edge $(\treenode{\afrun, p}, \treenode{\terminatingLasso, p}) \in E$ with $\terminatingLasso \in S^{K}_{\baut[A]}$ and $\appear{\afrun}{\afrun_{l}} = K$.
\item[Otherwise:]
	we have that
	\begin{align*}
		& \sum_{\treenode{\afrun', p'}: (\treenode{\afrun, p}, \treenode{\afrun', p'}) \in E} p' \\
		& {} = \hphantom{{} + {}} \sum_{\treenode{\afrun a q, \frac{p \cdot (1 - \stopprob)}{\size{\amat(\afrun_{l})}}}: (\treenode{\afrun, p}, \treenode{\afrun a q, \frac{p \cdot (1 - \stopprob)}{\size{\amat(\afrun_{l})}}}) \land \afrun a q \in P \land \appear{\afrun}{\afrun_{l}} > 1} \frac{p \cdot (1 - \stopprob)}{\size{\amat(\afrun_{l})}}\\
		& \hphantom{{} = {}} + \sum_{\treenode{\terminatingLasso, p \cdot \stopprob}: (\treenode{\afrun, p}, \treenode{\terminatingLasso, p \cdot \stopprob}) \land \terminatingLasso \in S^{K}_{\baut[A]} \land \appear{\afrun}{\afrun_{l}} < K} p \cdot \stopprob \\
		& {} = \hphantom{{} + {}} \size{\amat(\afrun_{l})} \cdot \frac{p \cdot (1 - \stopprob)}{\size{\amat(\afrun_{l})}}\\
		& \hphantom{{} = {}} + \sum_{\treenode{\terminatingLasso, p \cdot \stopprob}: (\treenode{\afrun, p}, \treenode{\terminatingLasso, p \cdot \stopprob}) \land \terminatingLasso \in S^{K}_{\baut[A]} \land \appear{\afrun}{\afrun_{l}} < K} p \cdot \stopprob
		\intertext{since by definition of $E$, we have an edge $(\treenode{\afrun, p}, \treenode{\afrun a q, \frac{p \cdot (1 - \stopprob)}{\size{\amat(\afrun_{l})}}}) \in E$ for each $(\afrun_{l}, a, q) \in \amat(\afrun_{l})$}
		& {} = \hphantom{{} + {}} p \cdot (1 - \stopprob)\\
		& \hphantom{{} = {}} + p \cdot \stopprob \\
		& {} = p
	\end{align*}
	as required, since by definition of $E$ and $S^{K}_{\baut[A]}$, we have exactly one edge $(\treenode{\afrun, p}, \treenode{\terminatingLasso, p \cdot \stopprob}) \in E$ with $\terminatingLasso \in S^{K}_{\baut[A]}$ and $\appear{\afrun}{\afrun_{l}} < K$.
\end{description}
This concludes the proof that for each non-leaf node $\treenode{\afrun, p} \in \big(P \times (0, 1]\big) \cup \setnocond{\treenode{\emptyword, 1}}$, it holds that $\sum_{\treenode{\afrun', p'}: (\treenode{\afrun, p}, \treenode{\afrun', p'}) \in E} p' = p$, as well as the proof of Theorem~\ref{thm:discrete-distribution-inclusion}.

\end{document}